# TEM observations on palladium samples aged up to 8 years under tritium

M. Segard [a], E. Leroy [b], B. Evin [a], S. Thiébaut [a], A. Fabre [a], S. Challet [a], V. Paul-Boncour [b]

[a] CEA, DAM, Valduc, 21120 Is sur Tille, France
[b] Chimie Métallurgique des Terres Rares, Institut de Chimie et des Matériaux Paris-Est, CNRS, Université de Paris XII, UMR 7182, 2-8 rue Henri Dunant, F-94320 Thiais, France

**Abstract:** Transmission electron microscopy (TEM) experiments were carried out on aged under tritium palladium powder samples. The experiments were devoted to the study of the evolution of $^3$He bubbles (size, density) appearing during aging. The study, using a new sample preparation, was focused for the first time on long aging time, up to 8 years. The TEM results indicated the presence of nanometer size bubbles (2 - 3 nm in diameter) with densities close to $5 \times 10^{23}$ m$^{-3}$.

## 1 Introduction

Tritium is an essential element for future thermonuclear energy production. However, due to its high mobility and radioactivity, tritium handling and storage raise safety issues. The solution generally adopted is to store tritium as metal hydrides (then called 'tritides') with low equilibrium pressures.
Among the metals and compounds which reversibly absorb hydrogen, palladium is of particular interest because of its resistance to oxidation and poisoning, its fast kinetic of absorption and desorption, and its ability to retain the $^3$He generated by tritium decay in the matrix, by a self-trapping process [1], for at least 9 years [2]. Indeed, this property enables the recovery of an almost $^3$He-free gas.

The first experimental evidence of the existence of $^3$He bubbles in aged tritides was given by Thomas *et al* [3], using TEM, in a Pd sample aged for 2 months under tritium (He/Pd = 0.01). Bubbles measured 1.5 to 2 nm in diameter, and their density was about 5 to $10 \times 10^{23}$ bubbles /m$^3$. Further studies enhanced the observation of these nanometric bubbles in Pd samples aged under T$_2$ for 3 months [4], and ultimately for 8 months [5], showing no significant evolution of bubbles diameter or density during this small time span.
However, these studies were limited towards aging time because "bulk" Pd samples were used, *i.e.* disks of about 100 µm thick, thinned down before TEM observations. Indeed, aging under tritium longer than 8 months induced the fracture of disks. Thinning and consequently TEM observations were then impossible.

Recently, a new technique based on the use of Pd powder was developed. Therefore, there is no limitation towards the aging time. Thanks to this technique, we have been able to observe for the first time Pd samples aged for very long times under tritium, the older one being aged for 8 years (He/Pd = 0.27).
In this article, the experimental procedure will be first detailed. Of course, the technique developed for the preparation, from Pd powder, of the thin samples for TEM observations will be extensively described. The methods used for TEM observations and image analyses, to evidence the bubbles and calculate their size distribution and density, will also be detailed. Results will then be presented for Pd samples aged from 2.5 years (He/Pd = 0.10) to 8 years (He/Pd = 0.27) under T$_2$. These results will be discussed, considering the evolutions of bubbles densities and especially bubbles sizes for these long aging times. Observations will be compared to previous results on younger samples [5] and calculations from modeling [6].

## 2    Experimental procedure

### 2.1    Powder preparation

It should first be mentioned that samples used in this study were not initially dedicated to TEM observations. They partially came from previous works on the aging effects on Pd pressure–composition isotherms (PcT), mainly at 25 °C and 40 °C, during tritium storage.

Briefly, powder samples (mean particle size about 14 µm) were cleaned up by deuterium absorption –desorption cycles before tritium loading, and were aged at room temperature before PcT measurements [7].

As the initial tritium content in Pd may be different from one sample to another, the aging state was here characterized by the $^3$He content in the palladium samples, expressed as the He/Pd atomic ratio (instead of merely considering the aging time). Moreover, all Pd powders have not been aged in the same conditions: some samples called "replenished samples" have been maintained at a constant T/Pd ratio during aging time while the others "not replenished" have been subjected to a decrease of their T/Pd ratio with time due to radioactive decay of tritium.

The $^3$He content was first estimated from both the initial tritium stoichiometry and the aging time by applying the radioactive decay law (theoretical He/Pd, see table 1). It was finally measured on a small fraction of the samples using thermal desorption (samples located in a tubular furnace, with the quantity and composition of the released gases analyzed by volumetry and high-resolution mass spectrometry as a function of temperature).

It was also checked by calorimetry or by thermal desorption that tritium in Pd samples was as low as uncountable amount prior to TEM observations.

### 2.2    Thinning / TEM preparation

The samples were in the form of micronic powder too large to be directly observed in TEM. Suitable samples for TEM consist in thin sheets with a thickness less than 100 nm. To reach this specification, the samples were prepared by means of the ultra-microtomy technique. The aged Pd powders were embedded into cold epoxy resin (*Struers Epofix*) and thin slices of 70 nm were cut with a *Leica UltraCut* ultra-microtome and deposited onto copper grids. The sample cutting was performed at room temperature with a 30° diamond knife.

### 2.3    TEM experiments

The TEM observations were carried out at the ICMPE/CMTR laboratory in Thiais (France) on a *FEI Tecnai F20* operating at 200 kV. The samples were observed at room temperature. The $^3$He bubbles, considered as cavities inside the material, were imaged in bright field, applying a defocus serie: - 1 µm, 0 µm and +1 µm as demonstrated elsewhere [5]. The images were captured with a *Gatan Orius 1000* CCD camera operating with the *DigitalMicrograph* software. At least ten different regions were studied on every samples to obtain sufficient statistics for the results.

### 2.4    TEM images analysis

The *ImageJ* software was used for images analysis [8, 9]. For each region, a stack of images was created from the three recorded images. The image displacement appearing at focus changes was corrected by the *StackReg* plugin [10].

In the present work, the cavities detection has been undertaken with a semi-automatic process, assuming that cavities appear as bright or dark zones on the images, depending on the focus. But contrary to previous observations on bulk samples [4, 5], images obtained on powder samples do not have a continuous and uniform background (see fig. 1) which enables an easy bubbles detection. Indeed, there are big contrast variations probably due to high stresses in the starting material. Moreover, the amorphous nature of the resin used leads to an additional image contrast. Despite this, the bubbles are considered such as bright (dark) zones: this is interpreted on images like a higher (lower) grey level compared to surrounding one. Consequently, local variations around each image pixel have to be identified to detect bubbles. So, after the images alignment process, the cavities are detected by the *ImageJ* function "local threshold" on both the underfocused and overfocused images. If a cavity appears at the same location in both images, the measurement is recorded. Then, the bubble is segmented and different parameters are measured with the "Analyze particles" function of *ImageJ* (surface, Feret diameter [11]…).

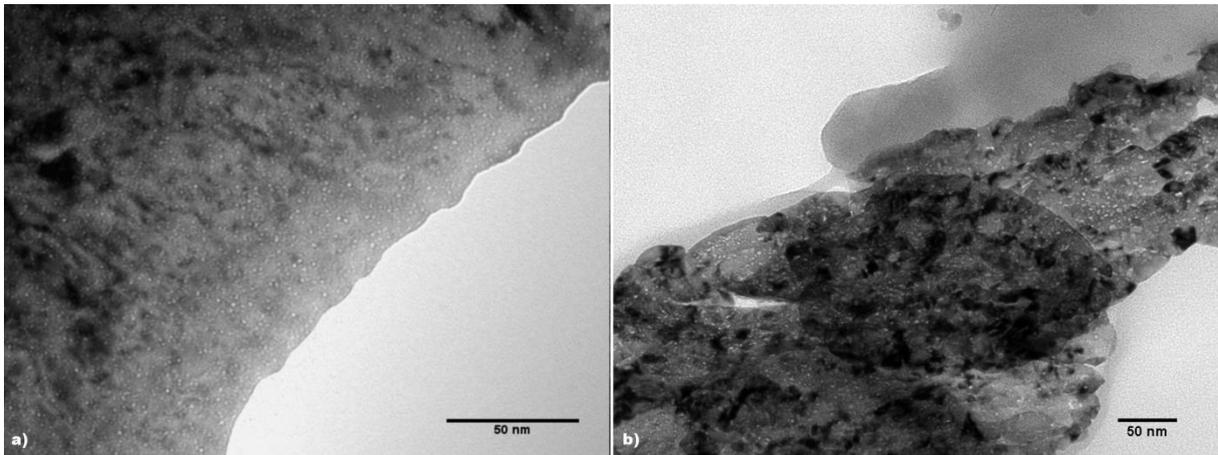

**Figure 1 : comparison between picture of bulk Pd sample coming from previous study (a) and picture of Pd powder sample embedded in resin from the present work (b).**

With these measurements, the size distribution of the $^3$He bubbles in each sample was plotted and its bubbles volumic density was calculated. Bubbles volumic density d, in [m$^{-3}$], corresponds to the number of bubbles on given area divided by the thickness of the ultra-microtomy slices:

$$d = \frac{N}{S \times e}$$

with N the number of characterized bubbles on a given surface S and a thickness e. In our case, the thickness of the sample was assumed to be 70 nm with an uncertainty of 10 nm, uncertainties on S and N are considered negligible. The bubbles density uncertainty is estimated to ± 1.10$^{23}$ m$^{-3}$ on all samples.

The bubble mean diameter $D_{bubble}$ is calculated working out the average of all measured bubble diameters. The uncertainty on the mean diameter is set as 10 %.

### 3   Experimental results

The figure 2 exhibits TEM underfocused and overfocused views of one of the youngest (He/Pd = 0.10, figure 2a) and of the two oldest samples (He/Pd = 0.23 and 0.27, figures 2b and 2c) presented in this study. The change of bubble size during the 8 years of aging can be directly observed.

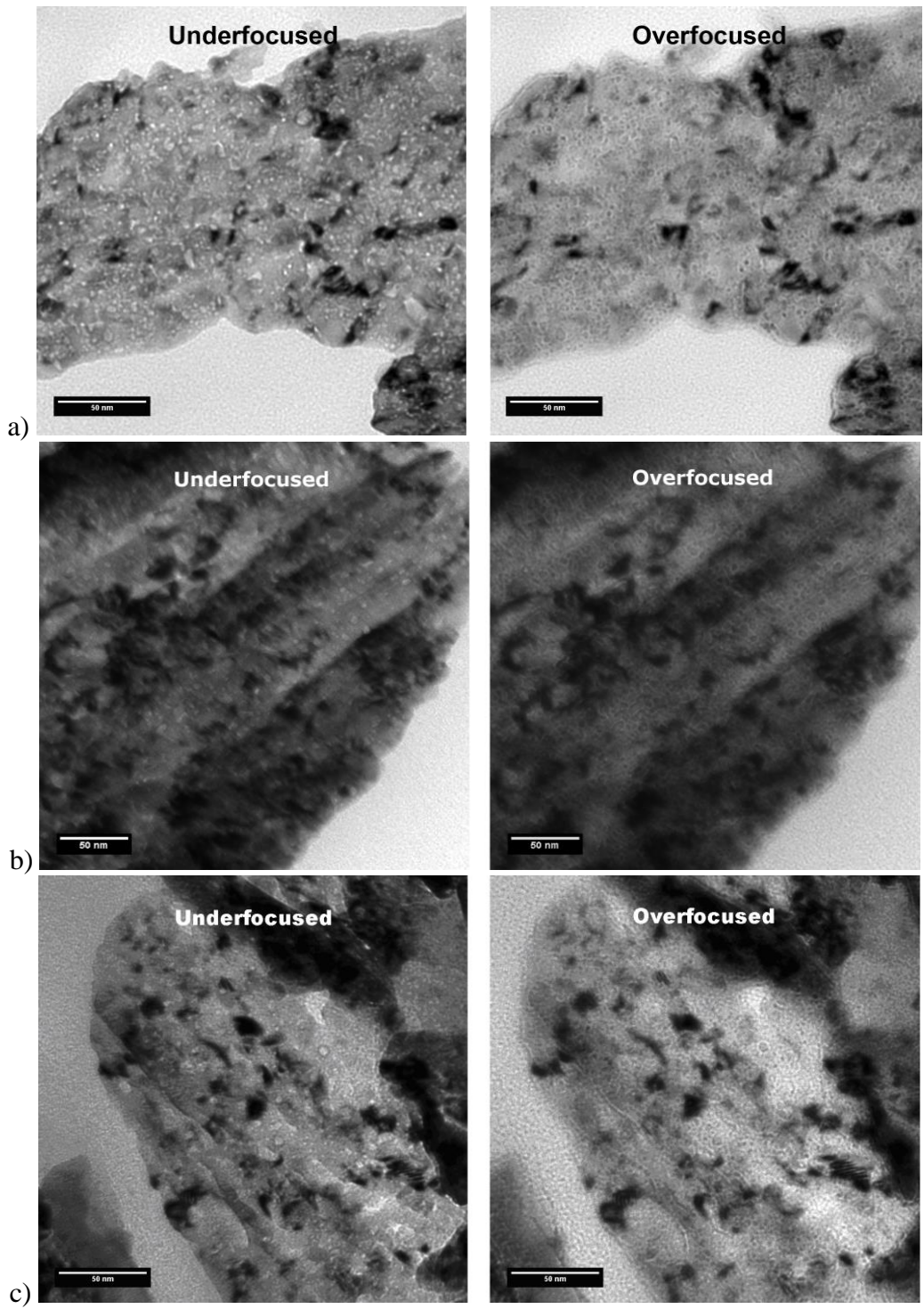

**Figure 2 : Focal series of He bubbles in (a) young Pd sample (He/Pd = 0.10) and in old samples (b: He/Pd = 0.23, c: He/Pd = 0.27).**

Bubbles characteristics coming from the TEM observations are gathered in table 1. For each sample, the mean bubble diameter (Feret diameter) and its standard deviation were calculated. The ranges of calculated bubbles densities are also indicated.

**Table 1 : bubbles characteristics measured in the aged Pd samples.**

| Theoretical He / Pd | Measured He / Pd | Aging conditions | Equivalent aging time [years] | number of bubbles for calculations | Bubble diameter | | | Bubbles densities |
|---|---|---|---|---|---|---|---|---|
| | | | | | Mean [nm] | Std dev [nm] | max. [nm] | [min. - max.] [x$10^{23}$ m$^{-3}$] |
| 0.10 | 0.10 | NR | 3.2 | 420 | 2.8 ± 0.3 | 2.1 | 19.1 | [1.7 - 4.3] ± 1 |
| 0.12 | 0.12 | NR | 3.4 | 1456 | 1.9 ± 0.2 | 1.1 | 9.8 | [1.3 - 6.9] ± 1 |
| 0.18 | 0.16 | NR | 5 | 2013 | 2.2 ± 0.2 | 1.3 | 15.5 | [2.0 - 11] ± 1 |
| 0.20 | 0.19 | NR | 6.3 | 300 | 2.3 ± 0.2 | 1.8 | 14.8 | [1.3 - 4.4] ± 1 |
| 0.23 | 0.21 | NR | 7.8 | 3965 | 1.9 ± 0.2 | 1.2 | 14.6 | [2.9 - 11] ± 1 |
| 0.23 | 0.23 | R | 6.7 | 3323 | 2.6 ± 0.3 | 0.7 | 7.8 | [0.4 - 70] ± 1 |
| 0.30 | 0.27 | R | 7.8 | 2451 | 2.8 ± 0.3 | 1.0 | 14.0 | [2.2 - 5.7] ± 1 |

**R = replenished sample - NR = not replenished sample**

## 4 Discussion

### 4.1 Bubbles sizes

First, these measurements indicate a good estimate of bubbles sizes during aging, which vary from 1.9 to 2.8 nm depending on the He/Pd ratio.

In comparison with previous studies [5], observed bubbles in the present work are bigger than those coming from younger samples in which highest bubble diameter was 1.5 nm (He/Pd = 0.02). The bubble growth is obviously highlighted even if it has to be moderated regarding the kind of sample (bulk sample used in [5] shaped as disks aged under tritium versus powder in this work) and regarding bubble densities (around $10^{25}$ m$^{-3}$ in [5]). Indeed, with similar aging conditions, higher bubbles densities lead to smaller bubble sizes.

The evolution of bubble radius against He/Pd is plotted in figure 3, in which is also added the measured bubble radii of younger samples coming from previous TEM observations of Thiébaut *et al.* [4] and Fabre *et al.*[5].

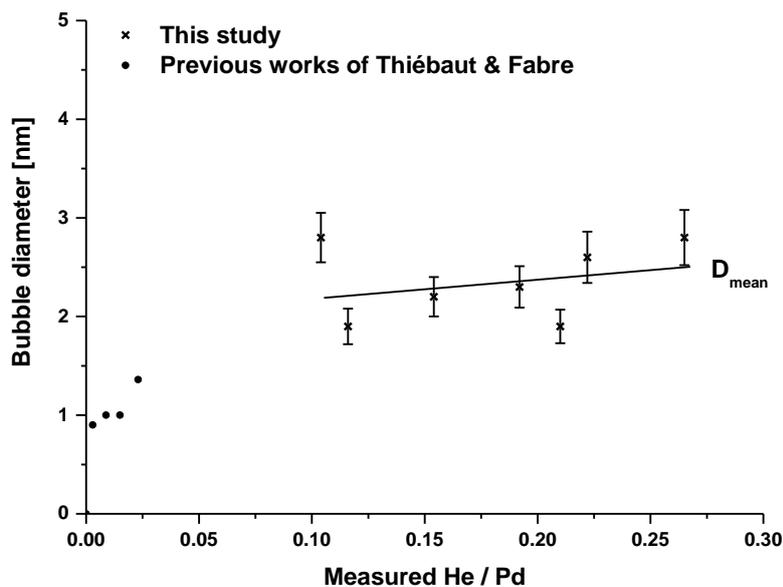

**Figure 3 : Evolution of the bubble diameter as a function of He/Pd.**

Standard deviations on bubble sizes are between 0.7 and 2.1 nm indicating fairly homogeneous distributions as visible in figure 4. Here the distributions look lognormal (straight line fit) on all samples.

The biggest bubbles observed range from 7.8 to 19.1 nm without dependence on aging time or aging condition. The smallest ones are approximately in the same range (≈ 0.6 nm) until He/Pd = 0.23 and then the smallest sizes increases up to 1.2 nm for the 2 oldest samples. The bubbles nucleation process never stops but this fact may indicate a decrease of the bubble nucleation rate. Nevertheless, as these samples were aged with replenished condition, this displacement can also be attributed to this. Indeed, in that case, the $^3$He generation rate from tritium decay is higher. It will be interesting in the future to compare these results to the ones coming from not replenished samples with same aging time.

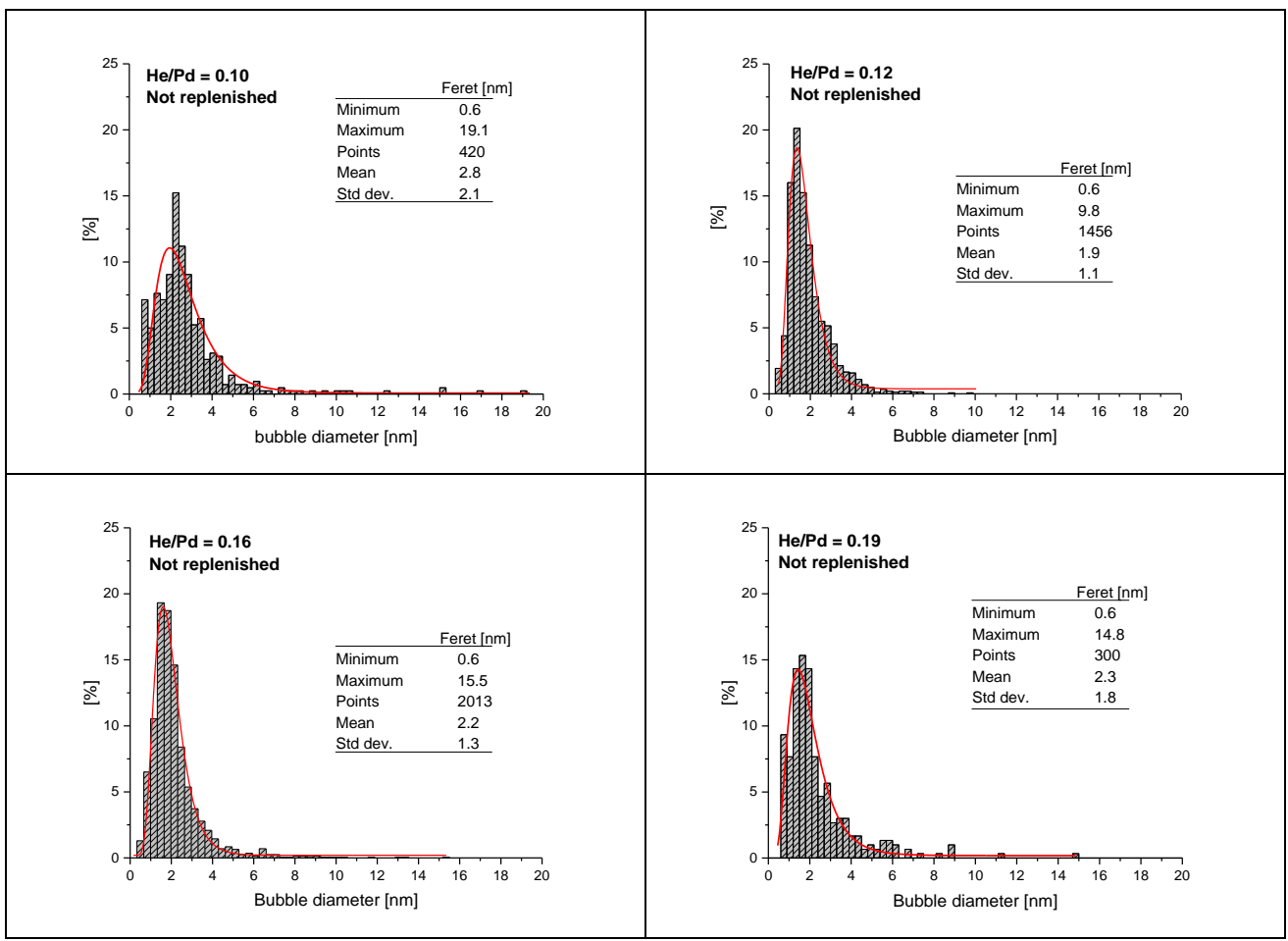

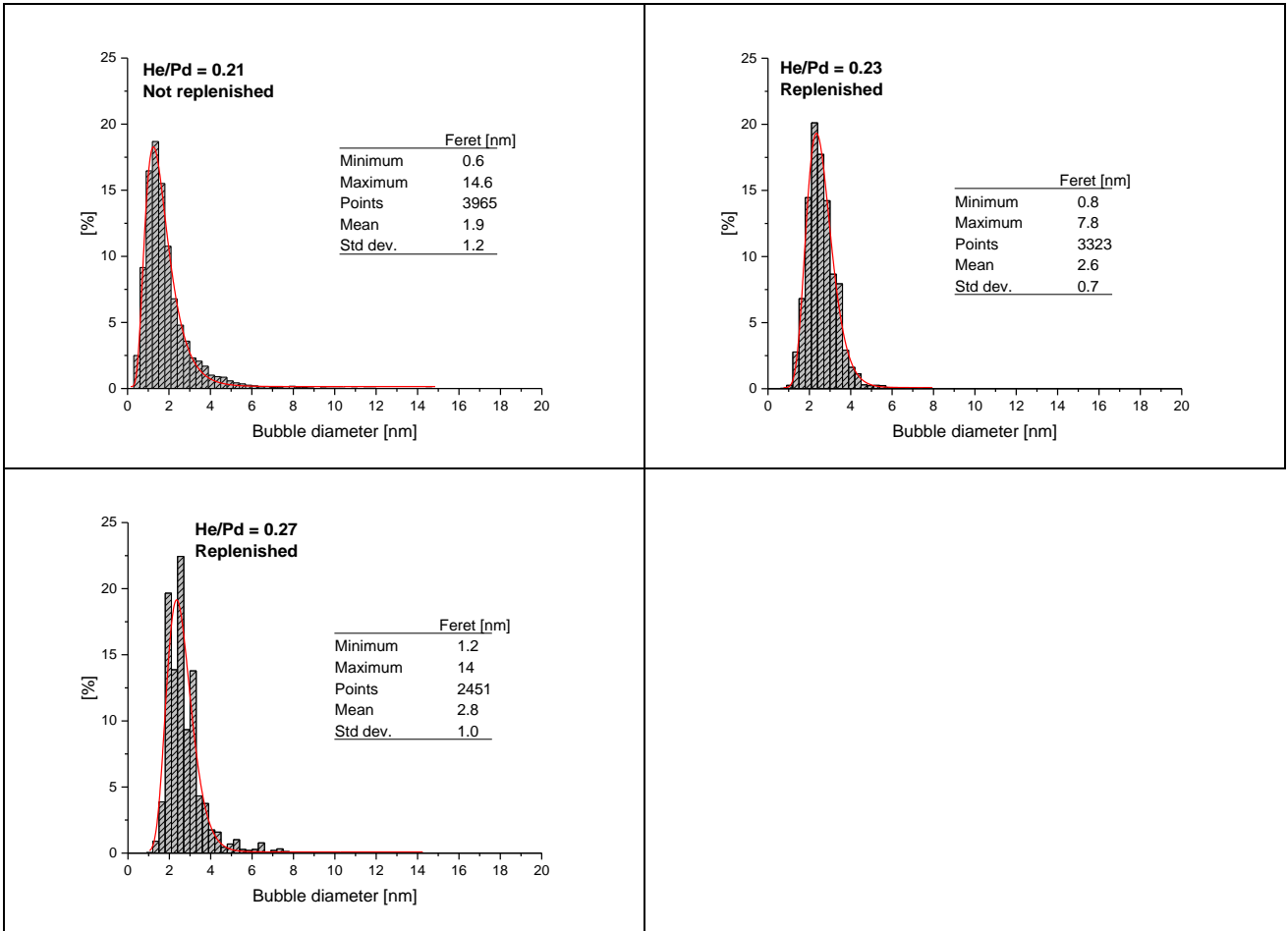

**Figure 4 : Sizes distributions of $^3$He bubbles in the aged Pd powder samples.**

### 4.2 Bubbles densities

Bubbles densities in these samples are mainly included between $1.10^{23}$ and $1.10^{24}$ m$^{-3}$. One of the studied areas of an old sample (He/Pd = 0.23) was characterized with a bubbles density of $7.10^{24}$ m$^{-3}$ and was excluded on figure 5. The range of bubbles densities is plotted as a function of He/Pd in figure 5. Even if the bubble nucleation process never stops, no significant evolution of the bubbles density regarding aging is observed, the magnitude of evolution of bubbles density is at the same level as uncertainties. This is consistent with predictions [6] indicating that bubbles generation in Pd tritide is established within the first weeks of aging. This also explains that aging conditions do not influence bubbles densities since, at this early aging time, the ΔT/Pd ratio between replenished samples and not ones is too small: the $^3$He generation rate and bubble nucleation rate are very similar in both cases.

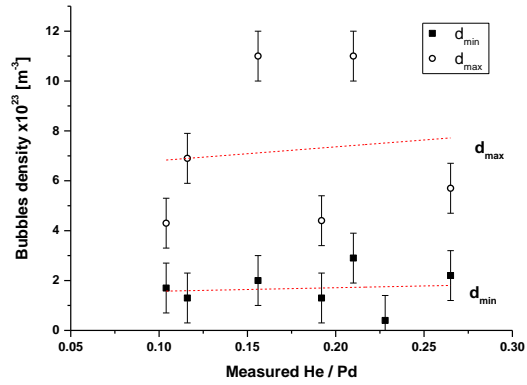

**Figure 5: Evolution of the range [$d_{min}$ ; $d_{max}$] of bubble densities as a function of He/Pd.**

In the present work, bubbles densities are quite lower than these measured on younger samples in previous works (see table 2). In fact, Thomas and Mintz have measured values close $1 \times 10^{24}$ m$^{-3}$ on samples aged 2 months (He/M = 0.006) whereas Thiébaut *et al.* have estimated bubbles densities between $3 \times 10^{24}$ and $1 \times 10^{25}$ m$^{-3}$ for samples aged from 1 month (He/Pd = 0.003) to 3 months (He/Pd = 0.01) [4]. In addition, Fabre *et al.* have evaluated densities between $5 \times 10^{24}$ and $2 \times 10^{25}$ m$^{-3}$ in 8-months-aged sample (He/Pd = 0.024) [5], that are ten-times-higher values than in the present work.

**Table 2 : Bubbles densities in aged Pd tritides found in literature.**

|  | **He/Pd (aging time)** | **bubbles density [m$^{-3}$]** |
|---|---|---|
| This study | 0.1 - 0.3 (3 - 8 years) | 1- 10.10$^{23}$ |
| Fabre *et al.* [5] | 0.024 (8 months) | 5.10$^{24}$ - 2.10$^{25}$ |
| Thiébaut *et al.* [4] | 0.003 - 0.01 (1 - 3 months) | 3.10$^{24}$ - 1.10$^{25}$ |
| Thomas and Mintz [3] | 0.006 (2 months) | 1.10$^{24}$ |

The difference between all studies can be explained by several assumptions. The first one is based on the shape of the aged Pd samples. Previous TEM studies have been made with bulk samples (disks) with a smaller surface/volume ratio than powders. Consequently the bubbles density was in that case reduced, as the $^3$He release fraction, at least within the first weeks of aging, was increased. In fact, when helium can be released out of the material, modeling shows that it has an effect on the concentration of isolated $^3$He atoms, $^3$He clusters formation and finally bubbles density is smaller. An example of $^3$He release rate effect on bubbles density is given in figure 6. Using the bubble nucleation model already mentioned [6] and applying arbitrary parameters, it can be noted that the higher the helium release rate, the smaller the bubbles density.

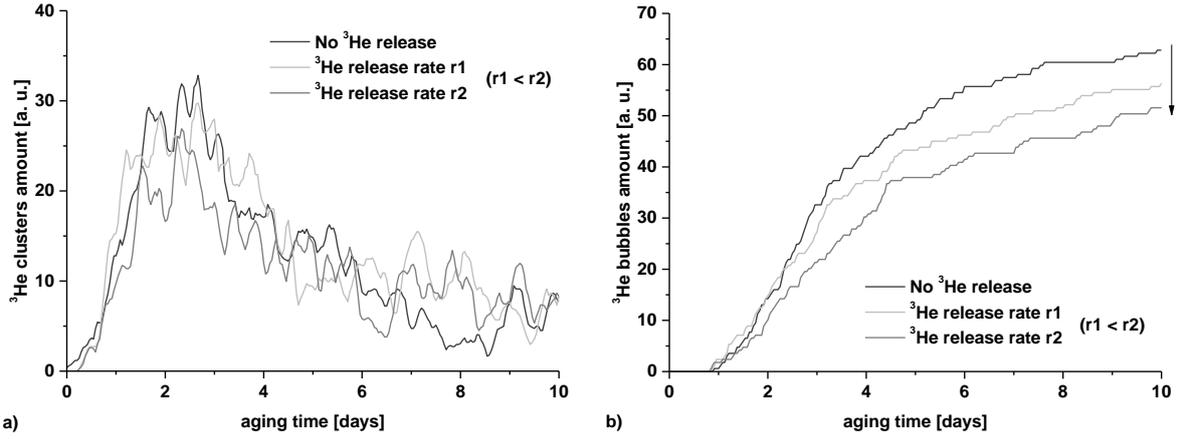

**Figure 6 :** a) Evolution of $^3$He clusters and b) $^3$He bubbles generated during the first days of aging as a function of $^3$He outgassing.

Secondly the amount of defects in the materials may also be involved to explain this discrepancy, since, theoretically, the presence of initial defects where He atoms can be trapped makes easier the bubble formation. Here, this assumption is hardly probable as palladium disks tend to "perfect" crystals (annealed 24h at 1000 °C prior to tritium charging) contrary to powders synthesized by precipitation of $PdCl_2$ and used as cast (XRD study on these powders has indicated defects initially present).

Finally, as discussed previously, the powdered Pd samples have been already used for other characterizations (several $\alpha \leftrightarrow \beta$ hydride phase transitions and slight heating below 50 °C) prior to TEM observations that may have induced $^3$He reorganization like bubbles coalescence.

### 4.3 Bubble growth modeling

To check the consistency between measured bubbles sizes and densities, a simple model giving the evolution of the bubble radius $R_{bubble}$ as a function of the He/Pd ratio and bubbles density has been developed for this study. It is assumed here, as a function of He/Pd, that there is no difference in bubble size evolution between replenished and non-replenished samples. Anyway, it is a time-dependent factor not introduced in that modeling.

First, as the He bubble pressure exceeds GPa values [12], the ideal gas law is inappropriate. For the pressure and temperature range of interest, a fit of the high pressure $^4$He EOS found by Le Toullec *et al.* at T = 298 K [13] is well suitable:

$$v = 23.810\, p^{-1/3} - 17.833\, p^{-2/3} + 29.760\, p^{-1} \quad (1)$$

where v is in [cm$^3$.mol$^{-1}$] and p is in [kbar]. It is assumed here the absence of $^3$He/$^4$He isotope effect.

A good fit of the inverted form of equation (1) can be given by:

$$\frac{1}{v} = C_1 \times p^{C_2} \quad (2)$$

with $C_1$ = 4.48 mol.m$^{-3}$.Pa$^{-C_2}$ and $C_2$ = 0.475. v is then expressed in [m$^3$.mol$^{-1}$] and p is in [Pa].

The term $\frac{1}{v}$ can be expressed as:

$$\frac{1}{v} = \frac{N_{He}}{V_{bubble}} \quad (3)$$

where $N_{He}$ is the bubbles $^3$He content in [mol] and $V_{bubble}$ is the volume of the helium bubbles in [m$^3$].

Combination of (2) and (3) leads to:

$$V_{bubble} = \frac{N_{He}}{C_1 \times p^{C_2}} = \frac{4}{3} \pi R_{bubble}^3 \quad (4)$$

At room temperature, the $^3$He bubbles in palladium tritide are supposed to grow by dislocation loop punching [14]. Trinkaus [15] relates the bubble pressure p to its radius $R_{bubble}$ by:

$$p = \frac{2\gamma + \mu b}{R_{bubble}} \quad (5)$$

where $\gamma$ is the surface free energy of the metal, $\mu$ is the shear modulus of the lattice and b is the Burgers vector of the dislocation loop.

Combination of (4) and (5) gives:

$$R_{bubble} = \left[ \frac{3 N_{He}}{4\pi \times C_1 \times (2\gamma + \mu b)^{C_2}} \right]^{\frac{1}{3-C_2}} \quad (6)$$

Assuming that all $^3$He atoms generated in the material are trapped into bubbles, the bubble $^3$He content $N_{He}$ is expressed as:

$$N_{He} = \left(\frac{He}{Pd}\right) \times \frac{N_{Pd}}{d_{bubble}} \quad (7)$$

$N_{Pd}$ is the amount of palladium by cubic meter in [mol.m$^{-3}$]; $d_{bubble}$ is the bubbles density in [m$^{-3}$].

Finally, the expression of $R_{bubble}$ is given by:

$$R_{bubble} = \left[ \frac{3\left(\frac{He}{Pd}\right) \times N_{Pd}}{4\pi \times C_1 \times (2\gamma + \mu b)^{C_2} \times d_{bubble}} \right]^{\frac{1}{3-C_2}} \quad (8)$$

The figure 7 shows the predictions of $R_{bubble}$ given by the model compared with $R_{bubble}$ measured by TEM. $\gamma$ and $\mu b$ were taken in literature [12]: $\gamma = 2$ N.m$^{-1}$ and $\mu b = 12$ N.m$^{-1}$.

Several bubbles densities were chosen to compute $R_{bubble}$. Indeed, as seen in fig. 7, using $d_{bubble} = 5.10^{23}$ m$^{-3}$ measured in this study, experimental results are not correctly fitted.
The most appropriate density for the bubbles sizes found by TEM is between $5.10^{24}$ and $1.10^{25}$ m$^{-3}$. The difference is greatly higher than the uncertainty on bubbles density ($1.10^{23}$ m$^{-3}$) detailed in the experimental procedure. Bubbles densities calculations from bubbles counting process are based on a constant sample thickness. It may be incorrect if the total sample thickness is composed of Pd and resin. In addition, the error in the bubble counting process is assumed negligible. As TEM pictures are projections of the material, some bubbles may be hidden by other ones, providing additional error in bubbles densities. Moreover, big contrasts in the TEM pictures make more difficult the bubbles counting.
Additional TEM observations combined with electronic tomography study, currently developing, will be useful to refine the actual results.

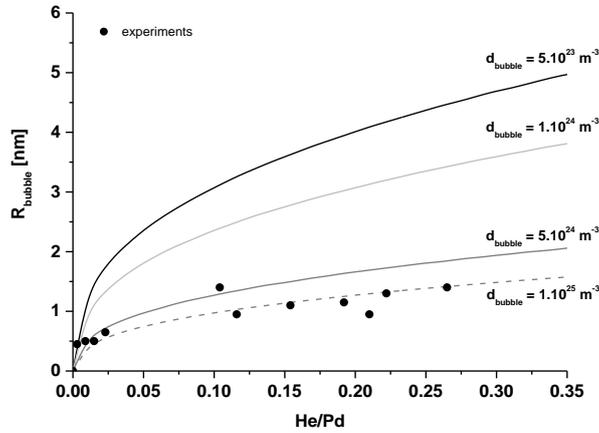

**Figure 7 : Bubble radius evolution as a function of He/Pd with different bubbles densities according to relation (8).**

## 5  Conclusion

In this work, palladium powder samples, aged under tritium up to 8 years have been observed by TEM for the first time. Bubble sizes between 1.9 and 2.8 nm in diameter have been measured, indicating a significant bubble growth regarding previous observations on younger samples (between 1.2 and 1.5 nm).
Measured bubbles densities are close to $5.10^{23}$ $m^{-3}$ in all samples (replenished or not), a value much smaller than in other studies ($\approx 10^{25}$ $m^{-3}$), which can be attributed to the nature (powder vs bulk), the initial state of our samples (defects, purity…) and/or errors in the bubbles counting process.
The evolution of the bubble size as a function of aging, characterized by He/Pd, has been compared to modeling. A good accordance has been found between model and experiments by applying densities between $5.10^{24}$ and $1.10^{25}$ $m^{-3}$.

To highlight influence of aging conditions on bubbles size evolution, future work will first be focused on TEM observations on dedicated samples (not used for other experiments such as PcT measurements), and all replenished to avoid α ↔ β phase transitions.
Observations of Pd powder samples aged some months under $T_2$ would be of great interest: a comparison of the results, especially the bubbles densities, with Pd disk aged the same time would permit to highlight the influence of the sample nature to this parameter.

Finally, with our technique developed for sample preparation (powder embedded in resin and ultra-microtomy cut), there is no limitation in aging time: the observation of samples which have reached the accelerated release stage is possible. This may help to identify phenomena (material fracture, $^3$He chemical potential equilibrium between bubble and solid solution…) at the origin of the ultimate stage of aging.

**Acknowledgement**

The authors gratefully thank C. Raddaz for the initial Pd sample preparation and V. Lalanne for the ultra-microtomy sample cutting.